\documentstyle[osa,manuscript]{revtex}

\begin{document}
\title{ONE- AND TWO-DIMENSIONAL SUBWAVELENGTH SOLITONS IN SATURABLE MEDIA}
\author{Boris V. Gisin}
\address{Department of Electrical Engineering - Physical Electronics,\\
Faculty of Engineering, Tel-Aviv University, Tel-Aviv 69978,\\
Israel; E-mail: gisin@eng.tau.ac.il}
\author{Boris A. Malomed}
\address{Department of Interdisciplinary Studies, Faculty of Engineering,\\
Tel Aviv University, Tel Aviv 69978, Israel; E-mail: malomed@eng.tau.ac.il}
\maketitle

\begin{abstract}
Very narrow spatial bright solitons in (1+1)D and (2+1)D versions of
cubic-quintic and full saturable models are studied, starting 
from the full
system of the Maxwell's equations, rather than 
from the paraxial (NLS)
approximation. For the solitons with both TE and TM polarizations, it is
shown that there always exists a finite minimum width, and they cease to
exist at a critical value of the propagation constant, at which their width
diverges. Full similarity of the results obtained for both nonlinearities
suggests that the same general conclusions apply to narrow solitons in any
non-Kerr model.
\end{abstract}

\section{Introduction}

The standard approach to description of spatial solitons in nonlinear
optical waveguides is based on the use of the nonlinear Schr\"{o}dinger
(NLS) equation, which replaces the Maxwell's equations (ME) in the paraxial
approximation \cite{books} (see also Refs. \cite{pre}). As is known, this
approximation can be insufficient for the description of very narrow, {\it %
subwavelength} (subwavelength) solitons, with a size $_{\sim }^{<}$ $\lambda 
$, where $\lambda $ is the carrier wavelength in vacuum \cite{Canberra}.
Terms which are neglected when deriving the NLS equation from ME couple to
the propagation constant, affecting dynamics of very narrow solitons with
arbitrary polarization \cite{Canberra}. Particular cases of the TM
(transverse-magnetic)- and TE (transverse--electric) polarized subwavelength
solitons in media with purely cubic (Kerr) nonlinearity were considered in
detail in Refs. \cite{Soreq,Taiwan}. It was found that, in the TM case, the
size of both bright and dark solitons cannot be essentially smaller than $%
\lambda /2$. In the TE case, the size of the dark soliton is also limited
from below, while the bright soliton may formally be arbitrarily narrow;
however, narrow TE solitons are subject to a strong instability.

The analysis performed in those works was formal in the sense that
subwavelength soliton solutions in the model with the Kerr nonlinearity
imply unrealistically large values of the nonlinear correction $\Delta n$ to
the refractive index, $\Delta n\,\,_{\sim }^{>}\,\,1$. As is known \cite
{books}, the quadratic (in the amplitude of the electromagnetic field)
correction $\Delta n$, which defines the Kerr nonlinearity, is, as a matter
of fact, only the first term of the expansion in powers of the squared
field. A detailed analysis shows that results of the formal consideration,
based on dropping the higher-order corrections and using the ensuing
truncated model in the range where the quadratic correction $\Delta n\,$is
allowed to be large, are {\em ambiguous}: they strongly depend on a
particular stage of the analysis at which the higher-order terms are
omitted. This ambiguity is especially conspicuous in the case of the TM
narrow solitons: depending on the choice of the truncation stage, one
arrives at a conclusion that the width of the bright TM soliton remains
limited from below, or may become arbitrarily small.

In fact, the physical problem of the existence of subwavelength solitons is
unresolvable within the framework of the Kerr model, and the only
possibility to obtain a definite result is to adopt a more realistic
nonlinearity, with {\em saturation} of $\Delta n$ in some form for strong
fields. While an exact form of the saturation in real materials is not
usually known, there are two commonly used models, viz., the saturable
nonlinearity proper \cite{saturable}, and the cubic-quintic (CQ) \cite
{cubicquintic,bullets} one. The former nonlinearity is realistic for alkali
metal vapors \cite{vapor}, and the latter nonlinearity, which combines
self-focusing cubic and self-{\em defocusing} quintic terms,\ is well
documented to describe nonlinear optical properties of the PTS crystal \cite
{PTS}. Moreover, formation of (2+1)D solitons in this material, consistent
with the predictions of the CQ model, was recently observed in the
experiment reported in Ref. \cite{Stegeman2}. We also note that the (2+1)D
and (3+1)D versions of the CQ model have recently attracted considerable
attention as they give rise to stable or almost stable multidimensional
(spatial or spatiotemporal) solitons \cite{bullets}.

Our objective is to consider subwavelength spatial solitons of both TM and
TE types {\em parallel} in the CQ and saturable models, in order to arrive
at general conclusions concerning their width and other fundamental
properties. Moreover, both (1+1)- and (2+1)-dimensional [(1+1)D and (2+1)D]
cases will be considered, in the latter case the spatial soliton being a
cylindrical beam. We will conclude that, in both models and for both
dimensions, there always exists a finite minimum value of the soliton's
width, and fundamental properties of the subwavelength solitons are fairly
similar in these two models. On the basis of these results, one may infer
that, in fact, qualitative results obtained in this work (first of all, the
existence of the minimum width) apply to narrow spatial solitons in {\em any}
non-Kerr model.

The rest of the paper is organized as follows. In section 2, we derive,
directly from ME, general equations that describe narrow spatial TM solitons
in both (1+1)D and (2+1)D geometries. In section 3, we consider a much
simpler TE case. In fact, for this case soliton solutions are well known (in
particular, they can be found in an exact form in the (1+1)D geometry). This
allows us to rigorously prove that the relative soliton's width, divided by
the carrier wavelength, cannot be smaller than a certain minimum value. In
section 4, combining analytical considerations and direct numerical
solutions, we find (1+1)D and (2+1)D fundamental TM solitons and arrive at
the conclusion that their width is also limited from below.

\section{Equations for the transverse-magnetic case}

We start the analysis with the most nontrivial case of the TM polarization.
In this case, irrespective of the particular nonlinearity, the {\em real}
(physical) vectorial electric ${\bf {\cal E}}$ and magnetic ${\bf {\cal H}}$
fields can be taken as 
\begin{equation}
{\cal H}_{y}=-H(x)\cos \Phi ,\;\;{\cal H}_{x}={\cal H}_{z}=0,  \label{1+1H}
\end{equation}
\begin{equation}
{\cal E}_{x}=E(x)\cos \Phi ,\;\;{\cal E}_{z}=E_{z}(x)\sin \Phi ,\;\;{\cal E}%
_{y}=0  \label{1+1E}
\end{equation}
in the (1+1)D case, and in the (2+1)D case, 
\begin{equation}
{\cal H}_{x}=(y/r)H(r)\cos \Phi ,\;\;{\cal H}_{y}=-(x/r)H(r)\cos \Phi ,\;\;%
{\cal H}_{z}=0,  \label{2+1H}
\end{equation}
\begin{equation}
{\cal E}_{x}=(x/r)E(r)\cos \Phi ,\;\;{\cal E}_{y}=(y/r)E(r)\cos \Phi ,\;\;%
{\cal E}_{z}=E_{z}(r)\sin \Phi ;  \label{2+1E}
\end{equation}
note a phase shift $\pi /2$ between the transverse and longitudinal
components in these expressions. Here, $x,y$ and $z$ are the transverse and
propagation coordinates, $r^{2}\equiv x^{2}+y^{2}$, $t$ is time, and the
common phase of all the fields is $\Phi =\beta z-\omega t$, $\beta $ and $%
\omega $ being, respectively, the propagation constant and frequency of the
carrier wave. In the (2+1)D case, the expressions (\ref{2+1H}) and (\ref
{1+1E}) correspond to the standard TM$_{n1}\,$mode in a cylindrical
waveguide \cite{Love}.

The simplest saturable model assumes an isotropic material characterized by
the relation 
\begin{equation}
{\bf {\cal D}}_{{\rm sat}}=\varepsilon _{0}{\bf {\cal E}}\left[ 1+\frac{%
\left( \varepsilon _{2}/\varepsilon _{0}\right) {\bf {\cal E}}^{2}}{1+\left(
\varepsilon _{4}/\varepsilon _{2}\right) {\bf {\cal E}}^{2}}\right]
\label{sat}
\end{equation}
between the electric induction and strength, with positive nonlinear
permeabilities $\varepsilon _{2}$ and $\varepsilon _{4}$. Expansion and
truncation of Eq. (\ref{sat}) yields the CQ model in the form 
\begin{equation}
{\bf {\cal D}}_{{\rm CQ}}={\bf {\cal E}\,}\left( \varepsilon
_{0}+\varepsilon _{2}{\bf {\cal E}}^{2}-\varepsilon _{4}{\bf {\cal E}}%
^{4}\right) \,.  \label{cuq}
\end{equation}

A difference between the two models appears in the case when the truncation
leading from ${\bf {\cal D}}_{{\rm sat}}$ to ${\bf {\cal D}}_{{\rm CQ}}$ is
no longer valid. Below, we present analysis in a detailed form for the CQ
model. For the saturable one, it is quite similar, but formulas are more
cumbersome. Final results will be displayed for both models together.

Following the usual rotating-wave approximation, the next step is to
substitute Eqs. (\ref{1+1H}) and (\ref{1+1E}) or (\ref{2+1H}) and (\ref{2+1E}%
) into the relation (\ref{cuq}) (or (\ref{sat}), in the case of the
saturable nonlinearity), and collect all the contributions to the
fundamental harmonics $\sin \Phi $ and $\cos \Phi $, neglecting higher-order
harmonics. This yields direct relations between the electric-field induction
and strength, 
\begin{equation}
{\cal D}_{x,y}=\varepsilon _{t}{\cal E}_{x,y},\;{\cal D}_{z}=\varepsilon _{l}%
{\cal E}_{z},  \label{D}
\end{equation}
where, in the case of the CQ nonlinearity, effective nonlinear
susceptibilities are 
\begin{equation}
\varepsilon _{t}\equiv \varepsilon _{0}+\frac{1}{4}\varepsilon
_{2}(3E^{2}+E_{z}^{2})-\frac{1}{8}\varepsilon
_{4}(5E^{4}+2E^{2}E_{z}^{2}+E_{z}^{4}),  \label{Dt}
\end{equation}
\begin{equation}
\varepsilon _{l}\equiv \varepsilon _{0}+\frac{1}{4}\varepsilon
_{2}(E^{2}+3E_{z}^{2})-\frac{1}{8}\varepsilon
_{4}(E^{4}+2E^{2}E_{z}^{2}+5E_{z}^{4}).  \label{Dl}
\end{equation}

Insertion of the above expressions (\ref{1+1H}) and (\ref{1+1E}) or (\ref
{2+1H}) for the magnetic and electric field and (\ref{2+1E}) into the
Maxwell's vectorial equation for the electric field,
\[
\nabla \times {\bf {\cal E}}=-\frac{1}{c}\frac{\partial {\bf {\cal H}}}{%
\partial t}\,,
\]
where $c$ is the light velocity in vacuum, we arrive at a scalar ODE
\begin{equation}
E_{z}^{\prime }+\beta E=-(\omega /c)H,  \label{E}
\end{equation}
where the prime stands for $d/dx$ or $d/dr$ in the (1+1)D and (2+1)D cases,
respectively. Further, we substitute the (1+1)D or (2+1)D expressions (\ref
{1+1H}) and (\ref{1+1E}) or (\ref{2+1H}) and (\ref{2+1E}), in combination
with the relations (\ref{D}) and (\ref{Dt}), (\ref{Dl}) which define the
electric induction, into the Maxwell's vectorial equation for the magnetic
field,
\[
\nabla \times {\bf {\cal H}}=\frac{1}{c}\frac{\partial {\bf {\cal D}}}{%
\partial t}\,.
\]
This yields two more equations, one of which is again an ODE, while the
other one is just an algebraic relation, 
\begin{eqnarray}
\varepsilon _{t}r^{1-D}(r^{D-1}H)^{\prime } &=&(\omega /c)E_{z}\varepsilon
_{l},  \label{HE} \\
\beta H &=&-(\omega /c)E,  \label{H}
\end{eqnarray}
where $D=1,2$ is the transverse dimension. 

Eliminating the magnetic field $H$ from Eqs. (\ref{E}) and (\ref{HE}) by
means of Eq. (\ref{H}), we then obtain a system of two equations for the
fields $E$ and $E_{z}$,  
\begin{eqnarray}
E_{z}^{\prime } &=&E(1-P),  \label{equ} \\
r^{1-D}\left( r^{D-1}Ev_{1}\right) ^{\prime } &=&-E_{z}v_{2}.
\label{equations}
\end{eqnarray}
Here, the variables have been rescaled as $(x,y,r)\rightarrow \beta \sqrt{%
\gamma }(x,y,r),\,\left[ E,E_{z}\right] \rightarrow \left[ E,(E_{z}\beta /%
\sqrt{\gamma })\right] (\omega /\beta \sqrt{\gamma })\sqrt{3\varepsilon _{2}}%
/2$, with $\gamma \equiv 1-\left( \beta _{0}/\beta \right) ^{2}$,$\;$ where $%
\beta _{0}\equiv \sqrt{\varepsilon _{0}}\omega /c$ is the propagation
constant in the linear regime, and 
\begin{eqnarray}
P &\equiv &E^{2}+\frac{\gamma }{3}E_{z}^{2}+\sigma \left( E^{4}+\frac{%
2\gamma }{5}E^{2}E_{z}^{2}+\frac{\gamma ^{2}}{5}E_{z}^{4}\right) ,  \label{P}
\\
v_{1} &\equiv &1+\frac{\gamma }{1-\gamma }P,  \label{v1} \\
v_{2} &\equiv &1+\frac{\gamma }{1-\gamma }\left[ \frac{1}{3}E^{2}+\gamma
E_{z}^{2}+\sigma \left( \frac{1}{5}E^{4}+\frac{2\gamma }{5}%
E^{2}E_{z}^{2}+\gamma ^{2}E_{z}^{4}\right) \right] ,  \label{nu2} \\
\sigma  &\equiv &-\gamma (1-\gamma )^{-1}s,  \label{sigma} \\
s &\equiv &\left( 10/9\right) \varepsilon _{0}\varepsilon _{4}\varepsilon
_{2}^{-2}.  \label{c}
\end{eqnarray}

These equations will be used below to study TE and TM solitons in the CQ
model, which will be paralleled by the same analysis for the model with the
saturable nonlinearity. Before proceeding to that, it may be relevant to
revisit the case of the TM soliton in the Kerr (rather than CQ) (1+1)D
medium. In that case, it is necessary to bear in mind the pure cubic (Kerr)
nonlinearity applies as long as the nonlinear correction to the refractive
index is much smaller than the linear index. In the present notation, this
condition can be shown to amount to inequalities
\begin{equation}
E^{2}(x)\ll 1{\rm \,}\text{and}{\rm \,\,}\left( E^{\prime }\right) ^{2}\ll 1.
\label{applicability}
\end{equation}

Taking these into regard and performing the corresponding expansions in the
above equations (\ref{equ}) through (\ref{nu2}), one can eliminate the
longitudinal electric field $E_{z}$ and derive an eventual equation for the
transverse field,
\begin{equation}
E^{\prime \prime }-\gamma E+\left[ E^{2}+\frac{1}{3}\left( E^{\prime
}\right) ^{2}\right] E=0\,,  \label{final}
\end{equation}
where $\gamma $ was defined by Eq. (\ref{gamma}). In comparison with the
traditional paraxial (NLS) approximation, the only new term in Eq. (\ref
{final}) is $\left( 1/3\right) \left( E^{\prime }\right) ^{2}E$, which, as a
matter of fact, is a contribution from the longitudinal component $E_{z}$ in
the present TM case.

Obviously, Eq. (\ref{final}) can give rise to bright solitons only in the
case $\gamma >0$, which will be assumed to hold hereafter. Note that the
classical broad solitons correspond to the case $\gamma \ll 1$. In that
case, the broad soliton with the first correction produced by the extra term
can be easily found:
\begin{equation}
E_{{\rm sol}}(x)=\sqrt{2\gamma }\left[ \left( 1+\frac{\gamma }{18}\right) 
{\rm sech}\left( \sqrt{\gamma }\xi \right) -\frac{\gamma }{9}{\rm sech}%
^{3}\left( \sqrt{\gamma }\xi \right) \right] \,.  \label{corrected}
\end{equation}

Numerical solution of Eq. (\ref{final}) shows that it does give rise to
solitary-wave solutions with an arbitrary small width; however, when the
soliton becomes too narrow, the solution violates the applicability
conditions (\ref{applicability}), which makes it necessary to modify the
nonlinearity, i.e., to consider the model with the CQ or saturable
nonlinearity, which will be done below.

\section{The transverse-electric case}

The above equations, derived for the CQ nonlinearity, also contain an
essentially simpler case of the TE polarization, which can be obtained
setting formally $\gamma =0$ but keeping $\sigma \neq 0$. In this case, $%
v_{1}=v_{2}=1$,$\;P=E^{2}+\sigma E^{4}$, and in the (1+1)D geometry an
eventual equation for $E(x)$ takes a well-known form 
\begin{equation}
E^{\prime \prime }=E-E^{3}-\sigma E^{5},  \label{1DCQ}
\end{equation}
a commonly known exact soliton solution to which can be written as 
\begin{equation}
4/E^{2}(x)=1+\sqrt{1+16\sigma /3}{\rm \cosh }(2x).  \label{1Dsolitons}
\end{equation}
It exists provided that $\sigma >-3/16$, or, with regard to Eq. (\ref{c}), 
\begin{equation}
(\beta /\beta _{0})^{2}<(\beta /\beta _{0})_{{\rm cr}}^{2}\equiv 1+3/(16s).
\label{upperlimit}
\end{equation}
As $(\beta /\beta _{0})^{2}$ is approaching the value $(\beta /\beta _{0})_{%
{\rm cr}}^{2}$, the average width of the soliton (\ref{1Dsolitons}), defined
as per Eq. (\ref{Waverage}), diverges as $\left| \ln \left[ 1+3/(16s)-(\beta
/\beta _{0})^{2}\right] \right| $, while the soliton's amplitude remains
finite, $E_{\max }=2$.

The main characteristic of the spatial soliton is its width. For the (1+1)D
soliton (\ref{1Dsolitons}), the FWHM \cite{books} width $\Delta $ can be
easily found, 
\[
{\rm \cosh \,}\Delta =\left( 1+16\sigma /3\right) ^{-1/2}+2. 
\]
In unnormalized units, the FWHM width is 
\[
W=(4\lambda /3\pi n_{0})\sqrt{s}F(16\sigma /3), 
\]
where 
\begin{equation}
F(y)\equiv \left| y\right| ^{-1/2}\ln \left[ (1+y)^{-1/2}+2+\sqrt{%
3+4(1+y)^{-1/2}+(1+y)^{-1}}\right] \,.  \label{F}
\end{equation}
The function $F$, and hence the width of the TE soliton, attain a minimum
value at $\sigma =-0.1493$. The ratio of the corresponding minimum width of
the TE soliton, normalized to the carrier wavelength, is 
\begin{equation}
W_{\min }^{{\rm (TE)}}/\lambda =0.825\sqrt{\varepsilon _{4}}/\varepsilon
_{2}\,.  \label{TEmin}
\end{equation}

In the next section, we will also use a different ({\it average}) definition
of the width, based on Eq. (\ref{Waverage}). It is easy to find that the
minimum value of the average width (\ref{Waverage}) differs from the minimum
FWHM value (\ref{TEmin}) by less than $8\%$.

Consideration of TE solitons in the (2+1)D geometry leads to an equation 
\begin{equation}
\frac{d^{2}E}{d^{2}r}+\frac{1}{r}\frac{dE}{dr}-\frac{1}{r^{2}}%
E=E-E^{3}-\sigma E^{5}  \label{2DCQ}
\end{equation}
($r$ again being the radial variable), cf. Eq. (\ref{1DCQ}). In fact, Eq. (%
\ref{2DCQ}) is exactly the same equation which describes known solitons with
an internal spin (vorticity) $s=1$ in the CQ model in the (2+1)D geometry 
\cite{bullets}. The origin of an effective ``vorticity'' term, $-E/r^{2}$,
in Eq. (\ref{2DCQ}) is the structure of the (2+1)D {\it ans\"{a}tze} (\ref
{2+1H}) and (\ref{2+1E}) for the vectorial fields.

Equation (\ref{2DCQ}) has been studied in detail numerically elsewhere \cite
{bullets}. Although the minimum width of a family of solitons generated by
this equation was not specially considered, a straightforward consequence of
results presented in Refs. \cite{bullets} is that the minimum width is
always finite.

Thus, it is possible to rigorously prove the existence of a finite lower
bound for the width of spatial TE solitons in the (1+1)D and (2+1)D models
with the CQ nonlinearity. Quite similarly, the same results can be obtained
for TE solitons in the model with the saturable nonlinearity (we do not
display technical details here, as they do not contain anything essentially
novel).

\section{Transverse-magnetic narrow spatial solitons}

The most interesting case is that with the TM polarization, as it does not
reduce to a single-component equation. As it was mentioned above, the
necessary condition for the existence of localized solutions to the
corresponding equations (\ref{equ}) and (\ref{equations}), supplemented by
Eqs. (\ref{P}) through (\ref{c}), is $\beta _{0}^{2}<\beta ^{2}$. Below, we
will use a {\it relative propagation constant}, $\beta /\beta _{0}\equiv
(1-\gamma )^{-1/2}$, as a measure of the departure from the paraxial
approximation: in the limit of a very broad soliton, one has $\beta /\beta
_{0}\rightarrow 1$, while in the opposite limit of an infinitely narrow
soliton (if any), $\beta /\beta _{0}\rightarrow \infty $.

A straightforward analysis of Eqs. (\ref{equ}) and (\ref{equations}) makes
it possible to prove that, in the (1+1)D case for both CQ and saturable
nonlinearities, localized solutions with no zeros of $E(x)$ and $E_{z}(x)$
at $x\neq 0$ do not exist if $E(0)=0$. Further analysis, details of which
are not displayed here, shows that, in the (1+1)D geometry, a localized
fundamental-soliton solution has an even monotonically decreasing transverse
component $E(|x|)$, while the longitudinal one $E_{z}(x)$ is odd, with $%
E_{z}(0)=0$ and a single extremum at finite $\left| x\right| $, see Fig. 1a
(in this and next figures, the results are simultaneously shown for the CQ
and saturable models). In the (2+1)D case, fundamental solitons feature a
different structure: they have $E(0)=0$, and both $E(r)$ and $E_{z}(r)$ may
have zeros at finite $r$, see Fig. 1b.

Proceeding to the width of the TM soliton, we use the electromagnetic energy
density, 
\begin{equation}
{\cal Q}={\cal E}_{x}{\cal D}_{x}+{\cal E}_{y}{\cal D}_{y}+{\cal E}_{z}{\cal %
D}_{z}+{\cal H}_{x}{\cal H}_{x}+{\cal H}_{y}{\cal H}_{y}\,,  \label{density}
\end{equation}
to define the soliton's average half-width as 
\begin{equation}
W=\int_{0}^{\infty }{\cal Q}r^{D}dr/\int_{0}^{\infty }{\cal Q}r^{D-1}dr.
\label{Waverage}
\end{equation}
The FWHM definition of $W$, which was used in the previous section for the
TE soliton, is ambiguous for TM solitons, as they have three different
components. The results for the average width, which were obtained from
numerical solutions of Eqs. (\ref{equ}) and (\ref{equations}), are
summarized in Fig. 2, showing $W$ vs. the relative propagation constant at
different values of the material constant $s$ defined by Eq. (\ref{c}).
Basic features of the dependence are the same as those following from the
above analytical expressions for the TE soliton in the (1+1)D case: there is
a {\em finite minimum} of the soliton's width, and the solitons do not exist
beyond a critical value of $\beta /\beta _{0}$, at which the soliton's width
diverges while its amplitude remains finite, cf. Eq. (\ref{upperlimit}).
Naturally, the minimum value of the width depends on the dimension of space,
type of the nonlinearity, and, for a fixed nonlinearity, on values of
material constants. As well as in the case of the TE solitons, the minimum
width tends to zero and the maximum value of $\beta /\beta _{0}$ diverges as 
$\varepsilon _{4}\rightarrow 0$.

Note that the soliton shapes shown in Fig. 1 pertain to the fixed value of
the relative propagation constant $\beta /\beta _{0}=1.05$, which is rather
close to the paraxial limit. Comparison with other numerical results
demonstrates that, for both CQ and saturable models, the shapes of the
(1+1)D and (2+1)D TM solitons remain fairly similar to those shown in Fig. 1
with the increase of $\beta /\beta _{0}$ up to the point where the minimum
width is attained. With the subsequent increase of $\beta /\beta _{0}$ up to
the value at which the solitons cease to exist, the shape changes much more.

The above considerations did not tackle the stability problem. While a
consistent stability analysis requires very tedious simulations of the
Maxwell's equations with the CQ or saturable nonlinearity, which is beyond
the scope of this work, we tried to test the stability by means of the
simple Vakhitov-Kolokolov (VK) criterion \cite{VK} (although its
applicability to the present model is not obvious): the propagation constant
must have a positive slope as a function of the beam's power (norm of the
solution), 
\[
N=2^{2-D}\left( 2\pi \right) ^{D-1}\int_{0}^{\infty }\left( {\cal E}_{x}^{2}+%
{\cal E}_{y}^{2}+{\cal E}_{z}^{2}\right) r^{D-1}dr 
\]
(cf. Eq. (\ref{Waverage})), which is a dynamical invariant of the model. Our
numerical calculations have shown that all the TM solitons, in both the
(1+1)D and (2+1)D geometries, satisfy the VK criterion.

\section{Conclusion}

In this work, we have studied very narrow (1+1)D and (2+1)D spatial TE\ and
TM solitons supported by the quintic-cubic and saturable nonlinearities,
starting the full Maxwell's equations. For TE solitons, we have obtained a
single equation, which is equivalent to the stationary version of the
nonlinear Schr\"{o}dinger equation with the corresponding nonlinearity,
while for TM solitons we end up with a system of equations for the
transverse and longitudinal components of the electric field, rather than
with a single one. It is also noteworthy that TE solitons in the (2+1)D case
are described by the same equation as vortex solitons with spin $s=1$ in the
(2+1)D nonlinear Schr\"{o}dinger equation.

Detailed analysis was displayed for the cubic-quintic model, while final
results were given for the saturable model as well. Full qualitative
similarity of the results obtained for both models (although exact scales
may be quite different, see caption to Fig. 2) makes it very plausible that
the situation is basically the same for {\em any} physically relevant
non-Kerr medium. The most important conclusions are that, in all the cases,
there is a finite minimum of the soliton's width, and the solitons cease to
exist at a critical value of the propagation constant, at which their width
diverges.

\newpage

\section*{Figure captions}

Fig. 1. Typical examples of the fundamental spatial TM soliton in the (1+1)D
(a) and (2+1)D (b) geometries. The relative propagation constant is $%
\beta/\beta _{0}=1.05$, and the material constant (\ref{c}) is $s=1$. In
this and next figures, the solid and dotted curves pertain, respectively, to
the cubic-quintic and saturable models.

Fig. 2. The width of the (1+1)D (a) and (2+1)D (b) fundamental spatial TM
solitons vs. the relative propagation constant $\beta /\beta _{0}$ at
various fixed values of the material constant $s$. For the saturable model,
the deviation of the relative propagation constant from $1$ is five times
that shown on the horizontal axis for the cubic-quintic model.

\end{document}